\documentstyle[12pt,aaspp4]{article}

\def\emfb{\overline{\mbox{\boldmath ${\cal E}$}} {}}
\def\emf{\overline{\mbox{${\cal E}$}} {}}
\def\bar{\overline}
\def\lint{\displaystyle \int}
\def\ren{R_{M}}

\def\bul{{\noindent $\bullet$\ }}
\def\part{\partial}

\def\bbA{\bar{\bf A}}

\def\bfE{{\bf E}}
\def\OB{{\overline B}}
\def\bfJ{{\bf J}}

\def\bfA{{\bf A}}
\def\bfB{{\bf B}}

\def\sm{{ M_\odot}}

\def\bbJ{\overline {\bf J}}

\def\bfa{{\bf a}}
\def\bfh{{\bf h}}

\def\lrp#1{\left(#1\right)}

\def\beqn{\begin{eqnarray}}
\def\eeqn{\end{eqnarray}}

\def\ni{\noindent}
\def\.{\mathaccent 95}
\def\beq{\begin{equation}}
\def\ee{\end{equation}}

\def\ep{\epsilon}

\def\frac#1#2{{\textstyle{{#1}\over {#2}}}}
\def\ni{\noindent}
\def\lsim{\mathrel{\rlap{\lower4pt\hbox{\hskip1pt$\sim$}}
    \raise1pt\hbox{$<$}}}
\def\gsim{\mathrel{\rlap{\lower4pt\hbox{\hskip1pt$\sim$}}
    \raise1pt\hbox{$>$}}}
\def\sqr#1#2{{\vcenter{\vbox{\hrule height.#2pt
         \hbox{\vrule width.#2pt height#1pt \kern#1pt
         \vrule width.#2pt}
         \hrule height.#2pt}}}}

\newbox\grsign \setbox\grsign=\hbox{$>$} \newdimen\grdimen \grdimen=\ht\grsign
\newbox\simlessbox \newbox\simgreatbox
\setbox\simgreatbox=\hbox{\raise.5ex\hbox{$>$}\llap
     {\lower.5ex\hbox{$\sim$}}}\ht1=\grdimen\dp1=0pt
\setbox\simlessbox=\hbox{\raise.5ex\hbox{$<$}\llap
     {\lower.5ex\hbox{$\sim$}}}\ht2=\grdimen\dp2=0pt

\def\doublespace {\smallskipamount=6pt plus2pt minus2pt
                  \medskipamount=12pt plus4pt minus4pt
                  \bigskipamount=24pt plus8pt minus8pt
                  \normalbaselineskip=24pt plus0pt minus0pt
                  \normallineskip=2pt
                  \normallineskiplimit=0pt
                  \jot=6pt
                  {\def\smallskip {\vskip\smallskipamount}}
                  {\def\medskip   {\vskip\medskipamount}}
                  {\def\bigskip   {\vskip\bigskipamount}}
                  {\setbox\strutbox=\hbox{\vrule 
                    height17.0pt depth7.0pt width 0pt}}
                  \parskip 12.0pt
                  \normalbaselines}

\font\gkvec=cmmib10                         
\def\bomega{\hbox{{\gkvec\char33}}}                  

\def\lb{\langle}
\def\rb{\rangle}
\def\bbE{\bar {\bf E}}
\def\bw{\bar{\omega}}
\def\bv{\bar V}
\def\bB{\bar B}
\def\ts{\times}
\def\lb{\langle}
\def\rb{\rangle}
\def\curl{\nabla {\ts}}

\def\bbV{\bar {\bf V}}
\def\bfv{{\bf v}}

\def\bfV{{\bf V}}
\def\bfj{{\bf j}}

\def\bfe{{\bf e}}
\def\bfw{{\bomega}}

\def\bfb{{\bf b}}

\def\bfB{{\bf B}}

\def\bbB{\bar{\bf B}}
\def\nb{\nabla}
\def\curl{\nb\ts}
\def\div{\nb\cdot}
\def\b0{b^{(0)}}
\def\v0{v^{(0)}}
\def\w0{\omega^{(0)}}
\def\bb0{\bfb^{(0)}}
\def\bv0{\bfv^{(0)}}
\def\bw0{\bfw^{(0)}}
\def\bj0{\bfj^{(0)}}

\def\ni{\noindent}
\begin{document}

\title{Distinguishing Propagation vs. Launch Physics 
of Astrophysical Jets and the Role of Experiments
} 
\author{Eric G. Blackman\altaffilmark{1}}
\affil{1. Dept. of Physics and Astronomy, Univ. of Rochester,
    Rochester, NY, 14627}

\begin{abstract}

The absence of other viable momentum sources  for collimated
flows leads to the likelihood that magnetic fields play a 
fundamental role in jet launch and/or collimation in astrophysical 
jets. To best understand the physics of jets,
it is useful to distinguish between  
the  launch region where the jet is accelerated 
and the larger scales where the jet propagates as a collimated
structure. Observations presently resolve jet propagation, 
but not the launch region. 
Simulations typically probe  the launch and propagation regions separately,
 but not both together. Here, I 
identify some of the physics of jet launch vs. propagation  and 
what laboratory jet experiments to date have probed.
Reproducing an astrophysical jet in the lab is unrealistic, so
maximizing the benefit 
of the experiments requires clarifying the astrophysical 
connection.

\end{abstract}

\section{Distinguishing Jet Launch vs. Jet Propagation Physics}

Jets in astrophysics emanate from accretion disk engines.
The available jet mechanical luminosity 
is inversely proportional to the radius from the central
engine, so the jet power is drawn from the inner most
regions of the disk. Material  must
be accelerated to outflow speeds comparable to
the escape speeds at the launch point.
Radiation pressure is typically incapable of providing the
directed  momentum and instead various flavors of 
magnetic launch models remain  the most promising.
(see Livio 2004, Pudritz 2004 for reviews; also Lynden-Bell 2003).

Magnetic  models take different forms.
In steady-state ``fling'' models (e.g. Blandford \& Payne 1982), mass flux
is sustained by centrifugal and toroidal magnetic pressure 
forces along the poloidal  field.  Explosive ``spring'' models
(e.g. Wheeler, Meier, \& Wilson 2002; Matt et al. 2004;
Moiseenko et al. 2006)
also thrive on a gradient of magnetic field pressure,
but  are time dependent and do not require
an initially imposed mass flux. Such ``springs'' 
may operate in gamma-ray bursts (GRB) and maybe supernovae.
In both spring and fling models, the launch region  
is Poynting flux dominated but on scales $\lsim 50R_{in}$,
(where $R_{in}$ is the scale of the inner engine)
the jet becomes flow dominated.

Springs and flings can be further distinguished from
magnetic tower  Poynting flux 
dominated outflow models 
(Lynden-Bell 2003; Uzdensky \& MacFadyen 2006);
for the latter, Poynting
flux domination remains even in the propagation region ($R\gsim 50R_{in}$).
Related models have been applied to GRB
and active galactic nuclei (AGN) assuming the baryon
loading is low. In the relativistic jets of  AGN, microquasars, and GRB it is not certain
how far in the propagation region the outflow remains PF dominated,
In the non-relativistic jets of  young stellar object (YSO), 
jets are baryon rich and likely flow dominated outside
the launch region. 

Presently, observations do not 
resolve the launch region at $R\lsim 50R_{in}$ for any source, 
although  best indirect evidence for MHD launch perhaps comes from 
rotation of YSO jets $\lsim100$ AU scales
(Coffey et al. 2004; Woitas et al. 2005).
That $B$-fields are important to jet  launching
($R\lsim 50 R_{in}$) is more widely agreed upon 
than the role of $B$- fields in the
asymptotic propagation region 
(despite the dearth of resolved observations of the former.)
For example, if, by $\sim 50R_{in}$, a magnetically 
collimated supersonic launch accelerates material to its asymptotic
directed supersonic speed, then the tangent of the opening angle
is just the inverse Mach number and the dynamical role of magnetic fields
at larger radii may be inconsequential.

In all standard magnetic jet models, the magnetic field is 
dominant in a corona above the rotator, 
and the magnetic field has a large scale, at least compared
to the scale of the turbulence in the rotator into which it  is anchored.
In recent magnetic tower models (Lynden-Bell 2003; Uzdensky
\& McFadyen 2006), the tower  has both
signs of vertical magnetic flux since it is composed of 
loops anchored with both footpoints in the engine.
Traditional MHD models which start with a large scale dipole field, 
produce a jet composed of one sign of magnetic flux and the return flux
meanders at large distances, being dynamically unimportant.
Instabilities in both geometries can disconnect blobs and produce knotty jets.

In short, the physics of the launch region (not yet resolved by astrophysical
telescopes) involves such issues as:
(1) Origin of magnetic fields, field buoyancy to coronae, 
magnetic helicity injection and relaxation into larger coronal structures,
(2) physics of centrifugal + magnetic 
acceleration of material from small to super-Alfv\'enic speeds,   
or Poynting flux driven bursts of acceleration,
(3) criteria for steady or  bursty jets, and 
(4) assessment of the extent of  Poynting flux domination.

The physics of the propagation region (resolved by astrophysical telescopes)
involves such issues as:
(1) Propagation, instability formation, and sustenance of collimation in 
as a function of  internal vs. external density and  
strength of magnetic fields, 
(2) bow shocks, cocoon physics, particle acceleration, 
(3) effect of cooling on  morphology, and 
(4) interaction with ambient media, stars, or cross-winds.

\section{Insights  on Launch from the Sun and a 
Two-Stage Paradigm for Jet Fields} 

Coronal holes and the solar wind
provide an analogy to the more extreme jet launching from
accretion disks. 
The launch region of the solar wind IS resolved.  
The coronal magnetic carpet (e.g. Schrijver \& Zwaan 2000) is 
composed  of large scale ``open'' field lines as well as smaller scale 
``closed'' loops. Both  reverse every 11 years, 
so we know that the field is not a residue
of flux freezing and must be produced by a dynamo.  

There are  three types of dynamos in astrophysics (e.g. Blackman \& Ji 2006):
(1) Velocity driven small scale dynamos, for which 
magnetic energy amplification occurs without sustained large scale flux
on spatial or temporal scales larger than the largest scale of the
turbulence,
(2) velocity driven large scale  dynamos which can amplify field 
 on {scales larger} than the largest turbulent scale, and  
(3) magnetically dominated large scale dynamos, 
also known as magnetic relaxation, 
whereby an already strong field, adjusts its geometry and 
such that any twists migrate to large scales.
Both type (2) and type (3) involve magnetic helicity and an 
associated  mean turbulent electromotive
force aligned with the local  mean magnetic field.

Type 1 and type 2 operate in the interior of a rotator, but 
some version of type 2, followed by
a type 3 dynamo,  provides the observable coronal field of the sun:
First, a velocity driven helical dynamo amplifies
fields of large enough scale that they buoyantly rise to the corona
without shredding from turbulent diffusion.  Once in the magnetically
dominated corona,
continued footpoint motions twist the field and inject magnetic
helicity. 
In response, the loops incur instabilities which
open up them  or make them rise. 
Fields that power jets from disks  may arise similarly.

The sun and disk are helicity injecting boundaries to their magnetically
dominated corona, (analogous to spheromak helicity injection (Bellan 2000)).
The type 2  dynamo occurs beneath the launch region
and  type 3 occurs in the  launch region.
 Neither occurs in the propagation region.

\section{Insights on Propagation and Launch 
from Jet Experiments}

Astrophysical jet experiments are in their first incarnation, 
and  presently involve  non-relativistic jet motion.
We cannot expect any experiment to reproduce 
any astrophysical source, but rather, address specific physics pieces.
To gain insight on astrophysical problems,  
a careful assessment of how a given experiment 
specifically relates to the Sec. 1 distinction between jet
formation and  propagation is required.

\subsection{Insights from Coaxial Gun Helicity Injection Experiments: Launch}

Hsu \& Bellan (2002)
employ a coaxial plasma gun composed of two coaxial electrodes
linked by an axisymmetric vacuum magnetic field.
This is analogous to an accretion disk with a dense set of 
poloidal magnetic loops, axisymmetrically distributed with zero
initial toroidal field.  At eight azimuthal locations, 
plasma is injected onto to the field
lines while an electric potential is driven across the anchoring
electrodes. An $\bfE\ts\bfB$ toroidal rotation of the plasma
results which then  twists
the poloidal field,  amplifying a toroidal component.
Equivalently, magnetic helicity is  injected along the field.

Once the twist is injected and the toroidal field amplified, 
the loops rise and merge on the axis.
(This is related to a type 3 dynamo, defined above.) 
A twisted unipolar core tower forms, rises, 
and remains collimated by hoop stress. The force free parameter
$\alpha_{inj}\equiv {\bf J}\cdot{\bf B}/B^2= I/\psi$ 
(where $I$ is the current from the imposed voltage across
the electrodes and $\psi$ is the initial poloidal
magnetic flux) measures
the amount of twist  injected. 
The  measurements roughly  agree with theoretical expectations
of the Kruskal-Shafranov kink instability criterion. 
For  $\alpha_{inj}\le 4\pi/L$, where $L$ is the magnetic column length, 
the collimated structure is stable.
When $\alpha_{inj}\gsim 4\pi/L$, the magnetic tower forms
exhibits a kink instability but the structure stays connected.
For $\alpha_{inj}>> 4\pi/L$, the magnetic tower forms,
a kink instability occurs, and a  disconnected magnetic blob 
forms.

The experiments show that a kink instability need not immediately
destroy jet collimation, even when disconnected blobs are produced. 
Real jets might be a series of ejected magnetically
blobs, rather than a continuous flow.
In astrophysics, pressure confinement may play an important
role of collimating any magnetic tower.

The experiments  probe jet formation in a plasma with 
$\beta \sim 0.02-0.1$,  $T\sim 5-20eV$,  fields of $\sim 
1$kG, and  number density $n=10^{14}$/cm$^3$.  
The Alfv\'en Mach number $<1$
so this is  a launch region experiment
not a propagation region experiment.
The value of $\alpha_{inj}$ in real astrophysical system is determined by shear,
resistivity, and coronal density.

\subsection{Insights from Pulsed  Wire Array Experiments: Launch and Propagation}


For the pulsed power machine MAGPIE, 
Lebedev et al. (2005) set up 
radial array of tungsten wires  arranged 
like spokes on a wheel (with a modest concavity) and applied
a radial current. The current ablates the wires and 
produces a magnetic field around each. The 
mean magnetic field has a net toroidal component in the plane above the 
array and  a net toroidal field of opposite sign below the array.
The  ${\bf J} \ts {\bf B}$ force from the toroidal magnetic pressure
accelerates material  vertically.
Toward the axis of the array, plasma is denser and an initial
hydrodynamic precursor jet forms. 

As the wires ablate and the magnetic force
accelerates the plasma, a magnetically dominated cavity forms,
Hoop stress
collimates the denser plasma along the axis of the tower. The axis becomes
a $\beta \sim 1$ plasma surrounded by $\beta <1$ toroidal magnetic field
dominated cavity. Outside of the cavity is a $\beta > 1$ ambient
plasma supplied from the early stage of wire ablation.
The dimensionless Reynolds and magnetic  Reynolds
numbers are larger than unity so the 
parameters are crudely OK for MHD an astro-comparison.
Cooling is important as the cooling length is short
compared to dynamical spatial scales.
The  experiment addresses  
principles of BOTH launch AND propagation physics.

The very narrowly collimated 
$\beta =1$ core jet has an internal Mach number of $\sim 4$.
The surrounding large toroidal magnetic pressure driven
cavity proceeds at Mach 10 with respect to the weakly magnetized 
ambient medium while the radial
expansion is only Mach 3 so the cavity is collimated.
In the experiments, these Mach numbers  are already reached even 
when the tower height is only of order  $\sim R_{in}$, where $R_{in}$
the array diameter $(\sim 4 mm.)$. 
That the vertical expansion is supersonic with respect to the
ambient medium implies that the jet head
has evolved from its formation region into its propagation region.

The structures produced in the experiments are  analogous to 
pressure confined  magnetic tower models (Lynden-Bell 2003; Uzdensky \& MacFadyen 2006), however there is very little polodial field, and the
net toroidal field is produced from poloidal loops oriented
perpendicular to the radial direction. 
A magnetically dominated tower encircling
a $\beta=1$ highly collimated core may also apply to astrophysical jets.

At later stages of the evolution, 
the magnetic tower becomes kink unstable and
a magnetic blob is ejected. But, as in Hsu \& Bellan (2002),
here too the instability does not destroy
the collimation of the tower. In this case, the ambient
thermal pressure slows radial expansion.
Blob formation again  highlights  the importance of  time dependent dynamics,
and that disconnected blobs may be the true nature of magnetized 
jets.  Were more material available from the wires, the blob 
formation process in the experiment could repeat.


Though not the main focus of Lebedev et al. (2005), it is important
to emphasize the precursor  jet which  precedes the magnetic cavity 
and results from the initially ablated plasma from the
inner region of the wire array. This jet is
hydrodynamic and collimated
by radiative cooling. In fact, the analogue of this precursor
jet is a close cousin to the 
the main focus of earlier conical wire array 
experiments of  Lebedev et al. (2002) and Lebedev (2004).
In these experiments, the conical array was more nearly 
cylindrical (concave at angles of 30 deg. with respect to the array 
axis rather than 80 deg.). Once the current is driven,
this lower inclination implies an increased density on the axis of the jet
compared to Lebedev (2005), thereby increasing 
cooling enough to break the flux freezing. 

Lebedev et al. (2002, 2004) are thus supersonic hydrodynamic jet 
experiments. Given the discussion of  Sec. 1,  
experiments for which the magnetic field is not important inside of jet 
are relevant at most to the propagation region, not the launch region.
The particular  hydrodynamic jet experiments do show that 
that collimated supersonic launch may obviate the need for asymptotic
magnetic collimation of a given jet when cooling is important.
The collimation is enhanced when the wire material has a larger
ion charge, enhancing radiative losses.
This is consistent with a model of asymptotic protostellar jet
collimation  discussed in  Tenorio-Tagle et al. (1988).


The Lebedev et al. (2002, 2004) experiments show Mach number $\ge 15$
jets. Jet deflection and shock propagation
are studied in Lebedev et al. (2004) experiments, where an additional
cross-wind
is introduced into across the propagating jet flow.
Generally, the hydrodynamic cooling-collimated 
jets seem to be relatively stable to non-axisymmetric perturbations.


\subsection{Insights from Laser Ablation Experiments:
Propagation} 

Another class of hydrodynamic 
jet experiments have been performed in laser inertial
confinement facilities (Blue et al. 2005; Foster et al. 2005).
These  probe aspects of the jet propagation regime only.
The experiments involve laser illumination of a 
thin metal disk such as titanium or aluminum.
The thin target is placed flush against a washer about 6 times thicker, 
often of the same material.
The  lasers ablate the thin target disk and the ablated
plasma is driven through the washer hole, exiting the hole
in the form of a supersonic jet.  The jet then propagates into
a foam. A variety of features can be studied from the jet propagation
into the foam using 
X-ray radiography and X-ray back-lighting.

Blue et al. (2005) report on experiments performed at NIF,
They studied  aspects of  nozzle 
angle on jet structure  by comparing  axially symmetric (2-D) vs. titled 
(3-D) nozzles.  The 3-D case leads to an earlier 
transition  
to turbulence than in the 2-D case.
Code testing of  2-D vs 3-D effects  
and the efficacy of the 3-D radiative HD code HYDRA (Marinak et al. 96)
was confirmed, although the Reynolds numbers of the experiment are 
$R=10^7$ while only $R=10^2-10^3$ in the simulations.


Similar  experiments performed by  Foster et al. (2005), on OMEGA 
obtain Mach numbers as high as $5$.
The images are somewhat clearer than in Blue et al (2005). 
Turbulent flows, dense plasma jets, bow shock structures
are seen. Modeling was done using 2-D hyrdo simulations with RAGE
(Gittings 1992). These experiments probe a  jet and foam density ratio of
$\rho_j /\rho_a \sim 1$. This is intermediate between 
YSO  jets which  have $\rho_j > \rho_a$ vs. 
AGN jets  which may have $\rho_j < \rho_a$.
The latter however, are relativistic, and the experiments
involve only non-relativistic flows.

\eject
\end{document}